\documentclass[runningheads]{svmult}

\usepackage{makeidx}   
\usepackage{graphicx}  
\usepackage{subeqnar}  
\usepackage{multicol}  
\usepackage{physprbb}  
\makeindex             

\begin{document}
\title*{A disk in the Galactic Center in the past?}
%
%
%
%
\titlerunning{A disk in the Galactic Center in the past?}
%
\author{Emmi Meyer-Hofmeister
\and Friedrich Meyer
\and Bifang Liu
}
\authorrunning{Emmi Meyer-Hofmeiste et al.}
%
%
\institute{Max-Planck-Institut f\"ur Astrophysik, Garching, Germany
}

\maketitle              

\begin{abstract}
We raise the question whether in the past a disk could have existed in 
our Galactic Center which has disappeared now. Our model for the 
interaction of a cool disk and a hot corona above (Liu et al. 2004) 
allows to estimate an upper limit for the mass 
that might have been present in a putative accretion 
disk after a last star forming event, but would now have evaporated by coronal action. 

\end{abstract}

\section{The evaporation process}

We study the possibility of a cool disk existing in the Galactic
Center in the framework of the disk-corona evaporation.
Interaction between the hot corona and the cool disk underneath occurs 
via energy and mass exchange. The hot corona conducts heat downward by 
electrons. At the bottom the heat is radiated away. 
An equilibrium is established: If the density in the corona is too
low, Bremsstrahlung is inefficient and the thermal conductive flux 
heats up some of the disk gas leading to mass evaporation from the disk 
into the corona. The resulting density increase in the corona
raises the radiation loss and thereby counteracts further
evaporation. If the coronal density is too high, radiative cooling is
too strong and gas condenses into the disk. 
At the final equilibrium density one of the processes works, either 
evaporation of disk mass to the corona or condensation of mass from 
the corona into the cool disk. The outcome depends on the mass flow 
in the corona from outside.

\section{The mass flow in the inner disk}

Assuming the applicability of our model for the evaporation/condensation 
process we can estimate how much mass would have evaporated by coronal 
action during a given time interval. We get an upper limit for the mass 
that might have been left over in a putative accretion disk after a last 
star forming event, assuming that no thin disk exists now.

We consider the region at the distance 
$10^4-10^5$  Schwarzschild radii from the black hole which corresponds to
$\frac{1}{400}$ to $\frac{1}{40}$ pc. For an assumed standard value of
the viscosity parameter $\alpha$ =0.3 and no
wind escape from the corona we found a rate of about $10^{-3}$ times the 
Eddington accretion rate corresponding to $10^{-4}$ solar masses per year. 

Chandra observations directly image the hot X-ray-emitting thermal gas in
the vicinity of the Bondi accretion radius where the surrounding gas
is captured by the gravitational pull of the central black hole,
and determine temperatures and densities that allow to estimate a mass
accretion rate of Sgr A$^*$  of $\dot M_{\rm{Bondi}}\sim (0.3-1)\times 
10^{-5}M_\odot$/yr (e.g. Baganoff et al. 2003).
If we assume that inflow rate during the time since the 
last star forming event, always disk evaporation would have been present.

\section{The history} 

We consider the history of our Galactic Center during the lifetime of the 
stars observed close to the Galactic Center. After the last star forming 
event in a then gravitationally unstable disk a certain amount of gas 
might have remained. This mass left over then should no longer have been 
gravitationally unstable. We here ask how much of this mass would have
evaporated. 

The observations, especially spectroscopy of one star, S0-2, 
observed in the vicinity of Sgr $\rm{A}^*$ suggest that these stars are main 
sequence O/B stars (Eisenhauer et al. 2003, Ghez et al. 2003). The 
main sequence lifetime of these stars is of order of $10^{6.5}-10^7$ years 
(Maeder \& Meynet 1989). We consider the
disk evolution during this time interval. From the evaporation rate we find 
an upper limit of 300 to 1000 solar masses for the amount of gas that
remained in a disk after these stars had been formed and was then
evaporated until now. 
This value is in the range of mass of the presently observed bright O/B 
stars close to the Galactic Center. This means about the same amount of mass 
as used up in star formation could have remained in a thin disk.

Such an estimate would also be of interest in the framework of star 
formation in the Galactic Center as recently discussed by Milosavljevi\'c 
and Loeb (2004). They suggest star formation in a warm
molecular disk where newly formed self-gravitating objects can have 
protostellar disks where fragmentation leads to multiple clumps
resembling the IRS 13 complex in the neighborhood of Sgr $\rm{A}^*$.

Interestingly this amount of gas is also close 
to that of the stability limit of a disk against self-gravitation. This 
suggests as a possible  picture for the evolutionary history that a disk 
could have become unstable by self-gravitation and formed the presently 
observed young massive stars around the Galactic Center until the 
gravitational instability had ceased. Even if such a disk could have
been too cool to allow magnetic dynamo action and would then have 
negligible internal viscosity and mass flow, it would have disappeared
by now by evaporation.

%

\end{document}